\documentclass[10pt,aps,twocolumn, prl, superscriptaddress, floatfix,showpacs]{revtex4-1}

\usepackage{graphicx}
\usepackage{amssymb}
\usepackage{amsmath}
\usepackage{bbm}
\usepackage{bm}
\usepackage{hyperref}
\hypersetup{breaklinks}
\usepackage{color}
\usepackage[normalem]{ulem}

\newcommand{\rb}[1]{\left( #1 \right)}
\newcommand{\ew}[1]{\langle #1 \rangle}
\newcommand{\beq}{\begin{eqnarray}}
\newcommand{\eeq}{\end{eqnarray}}
\newcommand{\eq}[1]{Eq.~(\ref{#1})}
\newcommand{\fig}[1]{Fig.~\ref{#1}}

\begin{document}
\title{Testing macroscopic realism through high-mass interferometry}
\author{Clive Emary}
\affiliation{
  Department of Physics and Mathematics,
  University of Hull,
  Kingston-upon-Hull,
  HU6 7RX,
  United Kingdom
}
\author{J. P. Cotter}
\author{Markus Arndt}
\affiliation{
University of Vienna,
Faculty of Physics,
VCQ \& QuNaBioS, 
Boltzmanngasse 5, 
A-1090 Vienna, Austria
}

\date{August 7, 2014}
\begin{abstract}
We define a quantum witness for high-mass matter-wave interferometers that allows us to test fundamental assumptions of macroscopic realism. 
We propose an experimental realisation using absorptive laser gratings and show that such systems can strongly violate a macrorealistic quantum-witness equality. 
The measurement of the witness can therefore provide clear evidence of physics beyond macrorealism for macromolecules and nanoparticles.
\end{abstract}

\maketitle

Are quantum superpositions of macroscopic objects possible, or does macroscopic realism --- the principle that a system exists in a  macroscopically-distinct state at all times\,\cite{Leggett1985,Leggett2002a,*Leggett2008} --- inevitably hold sway above a certain mass or size scale? 

Interferometry with massive objects provides a promising route to address this question and probe the macroscopic limits of quantum  coherence\,\cite{Arndt2014}.  Multi-slit diffraction has been demonstrated with molecules composed of more than $100$ atoms\,\cite{Juffmann2012a}. However, the collimation requirements in such experiments are stringent. Despite significant progress in the development of cold nano-particle sources in the mass range $10^{6}-10^{10}$\,AMU~\cite{Asenbaum2013,Kiesel2013,Bateman2014} it is Talbot-Lau interferometry that currently provides evidence for quantum interference using the most massive particles. Here, diffraction at each slit of a grating prepares the spatial coherence required to cover at least two neighbouring slits in a second mask. 
For gratings with the same period, separated by the Talbot distance $\xi_\mathrm{T}$, an interference pattern of equal period arises at a distance $\xi_\mathrm{T}$ further down stream. 
Talbot-Lau interferometers were first demonstrated using atoms\,\cite{Clauser1994a}, before being extended to hot molecules\,\cite{Brezger2002} and molecular clusters\,\cite{Haslinger2013}. Most recently quantum interference with molecules composed of more than 800\,atoms and a total mass exceeding $10^{4}$\,AMU has been observed\,\cite{Eibenberger2013}.

In practice, matter-wave experiments with high-mass particles require small grating periods because the de Broglie wavelength $\lambda_{dB} = h / m v$ is inversely proportional to the particle mass, $m$.  Here $v$ is the velocity and $h$ is Planck's constant.  If a grating is cut into a solid material, the small grating period result in high particle-surface interactions\cite{Juffmann2012a,Arndt1999}. In particular local patch potentials and charges on the surface are very difficult to control\,\cite{Grisenti1999,Brezger2003}. In contrast, gratings made from standing light waves\,\cite{Moskowitz1983,Gould1986} have well controlled particle interactions and both phase\cite{Gerlich2007} and absorption gratings\,\cite{Haslinger2013} have been adapted to Talbot-Lau interferometry. 

The interference patterns from the above experiments agree well with quantum-mechanical predictions, thus eliminating  classical ballistic models\,\cite{Nimmrichter2011}. 
However, the observation of interference itself does not rule out all possible macrorealist explanations.
Take, for example, the famous double-slit experiment for electrons\,\cite{Joensson1974b}.  Only by comparing the two-slit diffraction pattern with the sum of single-slit patterns is the quantum, wave-like, nature of the particles exposed\,\cite{Frabboni2008,Bach2013} and a macrorealist interpretation in terms of well-defined trajectories\,\cite{Robens2014} rejected.
In this Letter we propose a rigorous test to close this macrorealist loophole for high-mass particles in an all-optical Talbot-Lau interferometer.

\begin{figure*}[t]
\centering
    \includegraphics[width=1.8\columnwidth]{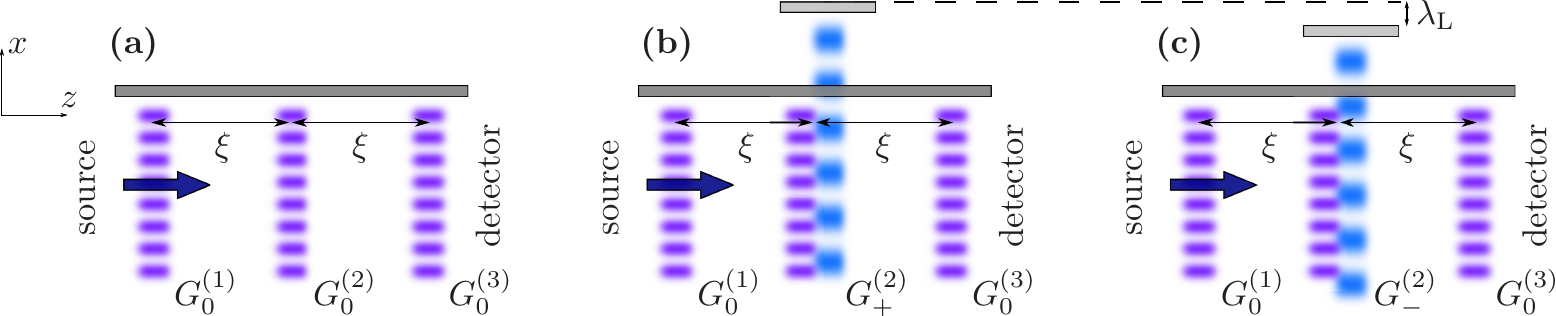}
    \caption{
      Talbot-Lau interferometer configurations for a quantum witness test.
      \textbf{(a)}:
      The interferometer consists of three absorptive standing-wave gratings, $G_0^{(i)};~i=1,2,3$, formed with lasers of wavelength $\lambda_\mathrm{L}$.
      \textbf{(b)}: 
      The middle grating is replaced by a combination of one laser beam with wavelength $\lambda_\mathrm{L}$ followed immediately, in position or time, by another with wavelength $2 \lambda_\mathrm{L}$. The second beam serves as a mask that blocks particles passing through the  ``even'' nodes of the original $G_0^{(2)}$ grating. We denote this combination $G_+^{(2)}$.
      \textbf{(c)}: The second beam is shifted by $\lambda_\mathrm{L}$ such that the ``odd'' nodes are blocked. We denote this combination $G_-^{(2)}$.
    \label{FIG:TLI}
    }
\end{figure*}

\fig{FIG:TLI}a illustrates an interferometer with three photo-depletion gratings, where the absorption of light removes particles that pass close to the antinodes. Gratings $G_0^{(1)}$, $G_0^{(2)}$ and $G_0^{(3)}$ are formed by the standing wave of three retro-reflected lasers of wavelength $\lambda_\mathrm{L}$. Absorptive optical gratings have been realized with continuous lasers for atoms~\cite{Abfalterer1997,Fray2004} and with pulsed lasers for molecular clusters~\cite{Haslinger2013,Doerre2014}. 
Figs.~\ref{FIG:TLI}b and \ref{FIG:TLI}c show the additional two experiments required for a test of macrorealism. Here the middle grating is supplemented by a second depletion laser, of wavelength $2\lambda_\mathrm{L}$, to form a grating that blocks every second node in $G^{(2)}$.  We designate the combined grating as $G^{(2)}_{+}$ when all ``even'' nodes are blocked and $G^{(2)}_{-}$ when the second laser is shifted to block all the ``odd'' nodes.

We characterise the difference in intensity distributions for the setting $G_0^{(2)}$ and the sum of the signals in the settings $G^{(2)}_{+}$ and $G^{(2)}_{-}$ with a {\em quantum witness}, $W$ \,\cite{Li2012a}. In analogy with entanglement witnesses\,\cite{Horodecki2009}, non-zero values of $W$ can be interpreted as witnessing quantum coherence. Whereas, if macrorealism holds, $W = 0$. The witness measurements should, from a macrorealist viewpoint, appear non-invasive. Therefore, switching from the one-colour to the two-colour setting should have no influence on the molecules other than blocking their paths. In practice realising non-invasive, ideal negative measurements\,\cite{Leggett1985, Emary2014} is extremely challenging. Here we quantify the invasivity of our measurement and find that it decreases with increasing values of $W$, enabling the witness to be attributed to a violation of macrorealism.

Within the eikonal approximation, a wave function $\psi(x)$ impinging on grating $G_0^{(i)}$ undergoes the transformation $\psi(x) \to t_0^{(i)}(x) \psi(x)$, where the transmission function\,\cite{Nimmrichter2011} of the three gatings in \fig{FIG:TLI}a is given by
\beq
  t_0^{(i)}(x) 
  = 
  \exp{\left[ 
    \left(
      - \frac{n_{0}^{(i)}}{2} + i \phi^{(i)}_{0} 
    \right)
    \cos^{2}{(k x)}
  \right]}
  ,
\label{EQ:t1}
\eeq
with $i=1,2,3$ and $k = 2\pi/\lambda_{L}$. Here, $n_{0}^{(i)}$ and $\phi_{0}^{(i)}$ are the mean number of absorbed photons and the dipole induced phase shift at the anti-nodes respectively. These are related to the absorption cross section $\sigma_\mathrm{abs}(\lambda_\mathrm{L})$ and  optical polarizabillity $\alpha(\lambda_\mathrm{L})$ through
$ 
  n_0^{(i)}/\phi_0^{(i)} 
  =  
  \beta(\lambda_\mathrm{L})
  =
 \lambda_\mathrm{L} \sigma_{\mathrm{abs}}(\lambda_\mathrm{L})/4\pi^2 \alpha(\lambda_\mathrm{L})
$.

\begin{figure}[tb]
\centering
    \includegraphics[width=0.9\columnwidth,clip=true]{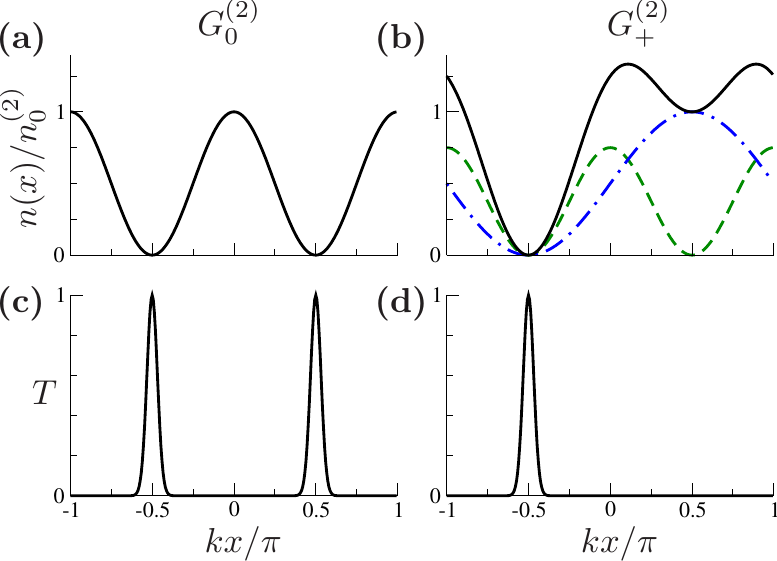}
    \caption{ 
      \textbf{(a)}   
      Mean number of photons $n(x)$ absorbed by a particle at position $x$ in the one-colour grating $G_0^{(2)}$.\textbf{(b)} The same for the two-colour grating $G_+^{(2)}$.
      Here, the total photon-number (black solid line) has contributions with period $\lambda_\mathrm{L}/2$ (green dashed) and $\lambda_\mathrm{L}$ (blue dot-dashed).  
      \textbf{(c)} and \textbf{(d)} show the corresponding transmission probabilities.
      With intensities chosen such that $n_0^{(2b)} = \frac{4}{3}n_0^{(2a)}=n_0^{(2)} \gg 1$, \eq{EQ:cond} is satisfied
      and grating $G_+^{(2)}$ has a transmission like that of  $G_0^{(2)}$ but with every other  ``hole'' closed.
    \label{FIG:int_n_T}
    }
\end{figure}

In \fig{FIG:TLI}b and c we show the complementary configurations created by adding a second absorptive laser of wavelength $2\lambda_\mathrm{L}$ to block every second opening of $G^{(2)}_0$.  The transmission functions of these composite gratings, $G^{(2)}_{\pm}$, read
\begin{align}
  t^{(2)}_{\pm}(x) 
  &=  
  \exp\left[ 
    \left(
      - \frac{n_{0}^{(2a)}}{2} + i \phi_{0}^{(2a)} 
    \right) 
    \cos^{2}{(k x)} 
  \right]
  \label{EQ:t2pm}
  \\ 
  &
  \times \exp\left[ 
    \left(
      - \frac{n_{0}^{(2b)}}{2} + i  \phi_{0}^{(2b)} 
    \right)
    \cos^{2}{\left(\frac{k x}{2} \mp \frac{\pi}{4}\right)} 
  \right]
  \nonumber
  .
\end{align}
with $\phi_{0}^{(2a)} =\beta(\lambda_\mathrm{L}) n_{0}^{(2a)}$ and $\phi_{0}^{(2b)} =\beta(2\lambda_\mathrm{L}) n_{0}^{(2b)}$.
\fig{FIG:int_n_T} shows the mean number of absorbed photons and the transmission probability at the second grating. The grating $G_0^{(2)}$ has openings in the unit cell at $x=\pm \lambda_\mathrm{L}/4$, whereas $G^{(2)}_{+}$ has a single opening at $x=- \lambda_\mathrm{L}/4$, and  $G^{(2)}_{-}$ an opening at $x=+ \lambda_\mathrm{L}/4$. In order to minimize the difference in electric field experienced by molecules that pass through $G^{(2)}_0$ and those that pass through $G^{(2)}_\pm$, we set
\beq
  n_0^{(2a)} = n_0^{(2)} 
  - 
  \textstyle{\frac{1}{4}}  n_0^{(2b)}
  \label{EQ:cond}
  .
\eeq
This means that for large photon absorption, $n_{0}^{(2)} \approx n_{0}^{(2b)} \gg 1$, we have to a good approximation
\beq
  |t_0^{(2)}(x)|^2 = |t_+^{(2)}(x)|^2 + |t_-^{(2)}(x)|^2
  \label{EQ:tbalance}
  .
\eeq

At the position of $G_0^{(3)}$ we define a dichotomic variable $Q$ such that those particles that pass through are assigned a value $Q=+1$, whilst those blocked are assigned a value $Q=-1$. The intensities recorded by a detector behind $G_0^{(3)}$ for the different grating settings of $G^{(2)}$ are  denoted by $I_{s_3,s_2}$.  Here $s_2=(0,+,-)$ refers to the three possible settings of $G^{(2)}$ and $s_3=(Y,N)$ describes the presence (Y) or absence (N) of the grating $G_0^{(3)}$ in the beam path. With grating $G_0^{(2)}$ in place, the fraction of molecules with $Q = +1$ can be obtained from the measured intensity ratio $I_{Y,0}/I_{N,0}$. The expectation value of $Q$ is therefore,
\beq
  \ew{Q} 
  = 
  2\frac{I_{Y,0}}{I_{N,0}} - 1
  \label{EQ:Q3}
  .
\eeq
Under a macrorealist, non-invasive description of the system the distribution of particles arriving at the detector when the second grating is in setting $G^{(2)}_0$ should equal the sum of intensities for the settings $G^{(2)}_{+}$ and $G^{(2)}_{-}$. We therefore define a second expectation value $\ew{Q}_\mathrm{m}$ which relates to settings $G_\pm^{(2)}$,
\beq
  \ew{Q}_\mathrm{m}
  = 
  2\frac{I_{Y,+} + I_{Y,-}}
        {I_{N,+} + I_{N,-}}
  -1
  \label{EQ:Q3m}
  .
\eeq
Finally, we combine $\ew{Q}$ and $\ew{Q}_\mathrm{m}$ to define the quantum witness\,\cite{Li2012a},
\beq
  W \equiv |\ew{Q} - \ew{Q}_\mathrm{m}|
  .
\eeq
For ideal non-invasive gratings, the condition $W = 0$ defines a macroreal state.  Finite values of $W$ describe the degree of quantum coherence present in a system with an algebraic bound $W \le W_{\mathrm{lim}} = 2$. We note that this is similar to the no-signalling in time measure\,\cite{Kofler2013}.

\begin{figure}[t!]
\centering
    \includegraphics[width=0.9\columnwidth]{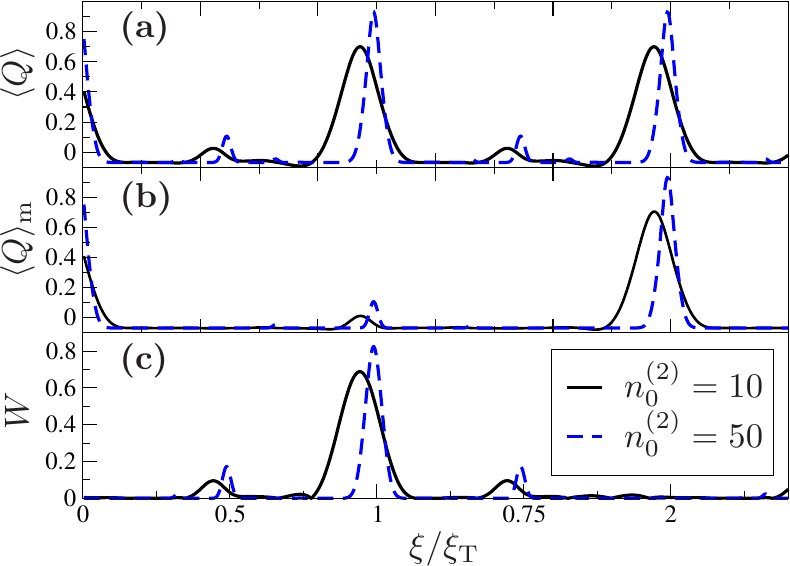}
    \caption{
      The expectation values $\ew{Q}$, $\ew{Q}_\mathrm{m}$, and quantum witness $W$ as a function of the ratio between grating separation $\xi$ and Talbot length $\xi_\mathrm{T} = \lambda_\mathrm{L}^2 /4\lambda_\mathrm{dB}$.
      The peaks in $\ew{Q}$ and $\ew{Q}_\mathrm{m}$ are a consequence of Talbot revivals. It is the wave nature of this phenomenon that results in differences between $\ew{Q}$ and $\ew{Q}_\mathrm{m}$.  As a consequence, the witness develops strong peaks at $\xi\approx (2q-1)\xi_\mathrm{T}$, which become sharper and taller with increasing $n^{(2)}_0$.
      The parameters used here are $n_0^{(3)} = 2$, $\beta(\lambda_\mathrm{L}) =\beta(2\lambda_\mathrm{L}) = 1$ and $n_0^{(2)}=10,50$.
    \label{FIG:QW}
    }
\end{figure}

\begin{figure}[t!]
\centering
    \includegraphics[width=0.9\columnwidth]{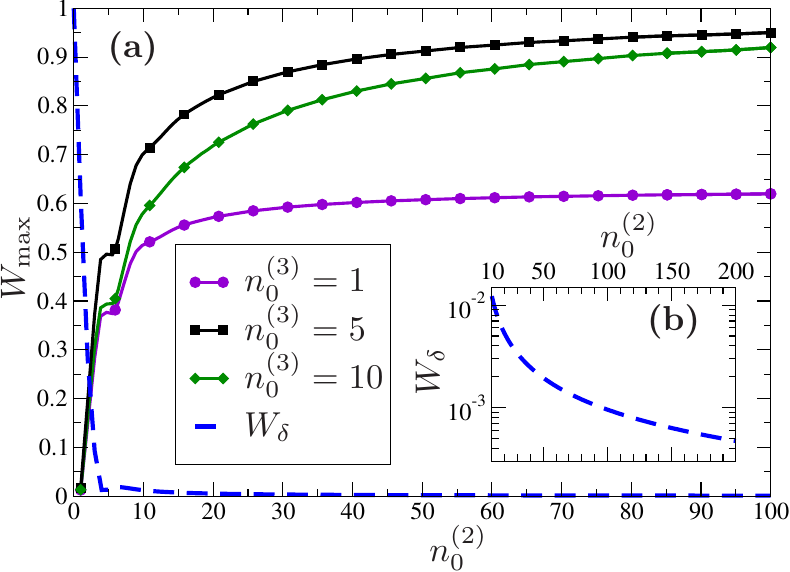}
    \caption{
      {\bf (a)}
      Maximum value of the quantum witness, $W_\mathrm{max}$, and  the revised upper bound, $W_\delta$, as a function of $n^{(2)}_0$ for $n_0^{(3)}=1,5,10$.
      {\bf (b)}
      The bound $W_\delta$ (log scale).  For $n_0^{(2)} \gtrsim 10$ this is several orders-of-magnitude smaller than the predicted violations.
      The parameters  here are the same as in \fig{FIG:QW}.
    \label{FIG:QWmax}
    }
\end{figure}

To calculate the quantum-mechanical value of $W$, we employ the Wigner-function method developed in Refs.~\cite{Hornberger2009,Nimmrichter2011a}. For simplicity, we set
$
  n_0^{(1)}
  = n_0^{(2b)}
  = \frac{4}{3}n_0^{(2a)}
  = n_0^{(2)}
$ throughout.
\fig{FIG:QW} shows the expectation values $\ew{Q}$ and $ \ew{Q}_\mathrm{m}$, and the witness $W$ for two different mean photon numbers.
Peaks in both $\ew{Q}$ and $\ew{Q}_\mathrm{m}$ arise due to Talbot revivals in the density distribution\,\cite{Berry1996}. 
For $\ew{Q}$, the ratio of grating periods is 1:1:1 and subsequent Talbot revivals appear when $\xi = q \xi_\mathrm{T}$,  where $q$ is a positive integer and $\xi_{T} = \lambda_\mathrm{L}^2 / 4 \lambda_\mathrm{dB}$ is the Talbot Length.
The interference pattern probed by $\ew{Q}_\mathrm{m}$ in \fig{FIG:QW}b arises from a setup where the ratio of the grating periods is 1:2:1 and in this configuration we expect the Talbot revivals at $\xi = 2q\xi_\mathrm{T}$.
Blocking half the slits in $G^{(2)}$ doubles the period of the intensity pattern at $G_0^{(3)}$, as expected ballistically. 
What differs from macrorealist expectations is that, rather than being shifted with respect to one another, the intensity patterns for gratings $G_+^{(2)}$ are $G_-^{(2)}$ are identical.
Thus, the peaks in $\ew{Q}$ at $\xi \approx (2q-1)\xi_\mathrm{T}$ 
have no analogue in $\ew{Q}_\mathrm{m}$, and this results in large non-zero values of the witness $W$ at those points. In \fig{FIG:QWmax}a we show the maximum value of the witness, $W_\mathrm{max}$, as a function of $n_{0}^{(2)}$. 
For optimum intensities, the maximum value approaches half the algebraic upper bound, $W_{\mathrm{max}} \approx W_{\mathrm{lim}}/2 = 1$.

One might be concerned that deviations from the exact equality in \eq{EQ:tbalance} can yield a macrorealistic explanation of finite values of $W$. These deviations, however, can be included in our analysis as a new macrorealistic upper bound for the witness. Let us define the difference in intensities $\delta_{s_3} \equiv I_{s_30}-I_{s_3+}-I_{s_3-}$.  If \eq{EQ:tbalance} holds, a macrorealist would concur that $\delta_{s_3} = 0$. However,  if \eq{EQ:tbalance} holds only approximately, they would conclude that the differences are finite because molecules can pass $G_0^{(2)}$ that would be blocked by the combination of $G^{(2)}_{+}$ and $G^{(2)}_{-}$.  Observing that $0 \le \delta_Y \le \delta_N$, a macrorealist would expect the witness to obey the revised bound 
\beq
  W \le W_\delta =  
    \frac{
      2 \left|\delta_N\right|
    }{
      I_{N,0}-\delta_N
    }
    .
\eeq
Here $W_{\delta}$ may be obtained experimentally from the position-integrated intensities measured without $G_0^{(3)}$. 
In \fig{FIG:QWmax}b we plot $W_{\delta}$ for a range of parameters and find that for $n_0^{(2)} \gtrsim 5$ it does not compromise the quantum witness violation.

A macrorealist would also argue that small differences between the gratings result in the witness being measured invasively. However,  the difference, and thus the potential for invasivity,  decreases with increasing laser power. For example, the difference in the number of photons absorbed by a molecule passing through $G^{(2)}_+$ compared to $G_0^{(2)}$ is 
$ 
  \Delta(x) 
  = 
  n_{0}^{(2)}
  \left[
    \cos^{2}{(kx/2 - \pi/4)} 
    -\frac{1}{4}
    \cos^{2}{(kx)}
  \right]
$. 
At the center of an opening in $G^{(2)}_+$, $\Delta(x)$ is zero but increases as $(x+\lambda_\mathrm{L}/4)^4$ close to this point. If we define the width of the opening by the points where the transmission has dropped to a value $T_w$, we find that the maximum difference experienced by particles traveling through the opening is 
$
  \Delta_\mathrm{max} \approx \rb{\ln T_w}^2/16 n_0^{(2)} \sim 1/n_0^{(2)}
$. 
Therefore, as the laser intensity increases the difference between the potentials experienced by the transmitted particles decreases and with it any invasivity. 
The increasing quantum witness violation for decreasing values of $\Delta_w$ makes an explanation for violations in terms of non-invasivity contrived.

A measurement of the quantum witness using a Talbot-Lau interferometer requires low-velocity particles which absorb sufficiently at $\lambda_{L}$ and $2\lambda_{L}$. In practice $G_0^{(1)}$ and $G_0^{(3)}$ may be replaced by material masks. However, $G^{(2)}_{0}$ and $G^{(2)}_{\pm}$ must be realised using absorptive optical gratings using photoionization\,\cite{Reiger2006}, photofragmentation\,\cite{Doerre2014} or any other technique which removes particles passing through the antinodes from detection. Caesium clusters with a mass exceeding $10^{5}$\,AMU can be created in great abundance using cold aggregation sources\,\cite{Goehlich1990} and can be ionised by radiation with a wavelength of $\lesssim539$\,nm\,\cite{Foster1969}.  Alkali metal clusters have an optical absorption cross section of approximately $10^{-20}$\,m$^{2}$ per atom\,\cite{Kreibig1995} which makes them easy to ionise in a $G^{(2)}_{\pm}$ formed from $266$\,nm and $532$\,nm continuous-wave lasers of modest power. Aggregation sources can also be adapted for use in a pulsed nanoparticle interferometer, operating in the time-domain, with velocities of less than $100$\,m/s\,\cite{Kousal2013}. Here, fluorine lasers with a wavelength of $\lambda_\mathrm{L}=157$\,nm can be combined with a frequency-doubled parametric oscillator to form $G_{\pm}^{(2)}$ with pulse energies in the mJ range.
In addition to metal clusters, this enables semiconductor nanocrystals and polypeptides and tryptophan clusters to be ionised\,\cite{Marksteiner2008,Marksteiner2009}.

In summary, we have presented a test of macrorealism for high-mass Talbot-Lau matter-wave interferometers. We have described an experiment that predicts significant values of the quantum witness and which can be realised in the near future using alkali metal clusters. This will enable the exclusion all macrorealistic, non-invasively-measurable theories for particles with a mass exceeding those of existing tests\,\cite{Robens2014} by several orders of magnitude\,\cite{Asadian2014}. 
Looking to the future, the development of single-photon depletion gratings for nano-biomatter\,\cite{Marksteiner2008} will extend interferometry to antibiotics, proteins and beyond.  Photo-activatable, mass-selected, fluorescent proteins are particularly appealing for this, as they can store which-path information in their internal structure. This will enable a post-selected measurement of the witness through real-time fluorescence detection
and allows the interplay between conformational state and decoherence to be explored\,\cite{Gring2010}.

We are grateful to N. D\"orre, S. Eibenberger, P. Haslinger, N. Lambert, M. Mayor, J. Rodewald, and U. Sezer for fruitful discussions. We thank the European Research Council project No. 320694 for financial support. JPC is supported by a VCQ fellowship.

\bibliographystyle{apsrev4-1}

\end{document}